\documentstyle[12pt,psfig]{article}
\begin{document}


\newcommand{\ep}{equivalence principle~}
\newcommand{\Ch}{Chandrasekhar~}
\newcommand{\Chp}{Chandrasekhar}
\newcommand{\Sc}{Schwarzschild~}
\newcommand{\Scp}{Schwarzschild}
\newcommand{\Sw}{Schwarzschild~}
\newcommand{\Swp}{Schwarzschild}
\newcommand{\Sch}{Schr{\"{o}}dinger~}
\newcommand{\Schp}{Schr{\"{o}}dinger}
\newcommand{\OVp}{Oppenheimer--Volkoff}
\newcommand{\OV}{Oppenheimer--Volkoff~}
\newcommand{\GR}{General Relativity~}
\newcommand{\GT}{General Theory of Relativity~}
\newcommand{\GRp}{General Relativity}
\newcommand{\GTp}{Reneral Theory of Relativity}
\newcommand{\STp}{Special Theory of Relativity}
\newcommand{\ST}{Special Theory of Relativity~}
\newcommand{\Lt}{Lorentz transformation~}
\newcommand{\Ltp}{Lorentz transformation}
\newcommand{\rel}{relativistic~}
\newcommand{\relp}{relativistic}
\newcommand{\msun}{M_{\odot}}
\newcommand{\eos}{equation of state~}
\newcommand{\eoss}{equations of state~}
\newcommand{\eossp}{equations of state}
\newcommand{\eosp}{equation of state}
\newcommand{\Eos}{Equation of state}
\newcommand{\Eosp}{Equation of state}
\newcommand{\beqn}{\begin{eqnarray}}
\newcommand{\eeqn}{\end{eqnarray}}
\newcommand{\nonum}{\nonumber \\}
\newcommand{\walecka}{$\sigma,\omega,\rho$~}
\newcommand{\waleckap}{$\sigma,\omega,\rho$}
\newcommand{\bbar}[1] {\mbox{$\overline{#1}$}} 
\newcommand{\courtesy}{~Reprinted with permission of Springer--Verlag 
New York; copyright 1997}
\newcommand{\oo}{{\"{o}}}
\newcommand{\au}{{\"{a}}}


\newcommand{\approxlt} {\mbox {$\stackrel{{\textstyle<}} {_\sim}$}}
\newcommand{\approxgt} {\mbox {$\stackrel{{\textstyle>}} {_\sim}$}}
\newcommand{\mearth} {\mbox {$M_\oplus$}}
\newcommand{\rearth} {\mbox {$R_\oplus$}}
\newcommand{\eo} {\mbox{$\epsilon_0$}}
\newcommand{\bfalpha} {\mbox{\mbox{\boldmath$\alpha$}}}
\newcommand{\bfgamma} {\mbox{\mbox{\boldmath$\gamma$}}}
\newcommand{\bfrho} {\mbox{\mbox{\boldmath$\rho$}}}
\newcommand{\bfsigma} {\mbox{\mbox{\boldmath$\sigma$}}}
\newcommand{\bftau} {\mbox{\mbox{\boldmath$\tau$}}}
\newcommand{\bfLambda} {\mbox{\mbox{\boldmath$\Lambda$}}}
\newcommand{\bfpi} {\mbox{\mbox{\boldmath$\pi$}}}
\newcommand{\bfomega} {\mbox{\mbox{\boldmath$\omega$}}}
\newcommand{\bp} {\mbox{\mbox{\boldmath$p$}}}
\newcommand{\br} {\mbox{\mbox{\boldmath$r$}}}
\newcommand{\bx} {\mbox{\mbox{\boldmath$x$}}}
\newcommand{\bv} {\mbox{\mbox{\boldmath$v$}}}
\newcommand{\bu} {\mbox{\mbox{\boldmath$u$}}}
\newcommand{\bk} {\mbox{\mbox{\boldmath$k$}}}
\newcommand{\bA} {\mbox{\mbox{\boldmath$A$}}}
\newcommand{\bB} {\mbox{\mbox{\boldmath$B$}}}
\newcommand{\bF} {\mbox{\mbox{\boldmath$F$}}}
\newcommand{\bI} {\mbox{\mbox{\boldmath$I$}}}
\newcommand{\bJ} {\mbox{\mbox{\boldmath$J$}}}
\newcommand{\bK} {\mbox{\mbox{\boldmath$K$}}}
\newcommand{\bP} {\mbox{\mbox{\boldmath$P$}}}
\newcommand{\bS} {\mbox{\mbox{\boldmath$S$}}}
\newcommand{\bdel} {\mbox{\mbox{\boldmath$\bigtriangledown$}}}

\newcommand{\eps} {\mbox {$\epsilon$}}
\newcommand{\gpercm} {\mbox {${\rm g}/{\rm cm}^{3}$}}
\newcommand{\rhon} {\mbox {$\rho_{0}$}}
\newcommand{\rhoc} {\mbox {$\rho_{c}$}}
\newcommand{\fmm} {\mbox {${\rm fm}^{-3}$}}
\newcommand{\bag} {\mbox {$B^{1/4}$}}
\newcommand{\fraca} {\mbox {$ \frac{1}{2}       $}}
\newcommand{\fracb} {\mbox {$ \frac{3}{2}       $}}
\newcommand{\fracc} {\mbox {$ \frac{1}{4}       $}}
\newcommand{\fraccc} {\mbox {$ \frac{1}{3}       $}}

\newcommand{\xn} {\mbox {$ x_{N}     $}}
\newcommand{\un} {\mbox {$ u_{N}    $}}
\newcommand{\vlambda} {\mbox {$ \frac{1}{4} \lambda (\sigma^{2} -
\sigma_{0}^{2} )^{2}      $}}
\newcommand{\kf} {\mbox {$ k_{F}       $}}
\newcommand{\stressenergy} {\mbox {$ 
\calT_{\mu \nu} = -g_{\mu \nu} \calL  + \sum_{\phi} \frac{\partial \calL}
{\partial (\partial^{\mu} \phi)} \partial_{\nu} \phi $}}
\newcommand{\mstar}[1] {\mbox {$ m^{\star}_{#1}      $}}
\newcommand{\ef}[2] {\mbox {$ (\kf^{2} + \mstar{#2}^{2})^{#1 /2}  $}}
\newcommand{\ek}[2] {\mbox {$ (k^{2} + \mstar{#2}^{2} )^{#1 /2}  $}}
\newcommand{\msigma} {\mbox {$ m_{\sigma}      $}}
\newcommand{\momega} {\mbox {$ m_{\omega}      $}}
\newcommand{\mrho} {\mbox {$ m_{\rho}      $}}
\newcommand{\mN} {\mbox {$ m_{N}      $}}
\newcommand{\mB} {\mbox {$ m_{B}      $}}
\newcommand{\munu} {\mbox {$ \mu \nu       $}}
\newcommand{\calT} {\mbox {$ {\cal T}       $}}
\newcommand{\calH} {\mbox {$ {\cal H}       $}}
\newcommand{\calL} {\mbox {$ {\cal L}       $}}
\newcommand{\calD} {\mbox {$ {\cal D}       $}}
\newcommand{\Lint}[1] {\mbox {$ \bar{\psi}_{#1} [ g_{\omega }
\gamma_{\mu} \omega^{\mu} + \frac{1}{2} g_{\rho }  \gamma_{\mu}
 \bftau\! \cdot\! \bfrho^{\mu} \cdots  ] \psi_{#1}  $}}
\newcommand{\LLint}[1] {\mbox {$ \bar{\psi}_{#1} [ - g_{\sigma}
\sigma + g_{\omega }
\gamma_{\mu} \omega^{\mu} + \frac{1}{2} g_{\rho }  \gamma_{\mu}
 \bftau\! \cdot\! \bfrho^{\mu} \cdots  ] \psi_{#1}  $}}
\newcommand{\Lsigma}[1] {\mbox {$ 
\bar{\psi }_{#1} [i \gamma_{\mu} \partial^{\mu} -g( \sigma
+i \gamma_{5}\bftau\! \cdot\! \bfpi) ] \psi_{#1}  $} }
\newcommand{\Lmeson} {\mbox {$
\frac{1}{2} (\partial_{\mu} \sigma \partial^{\mu} \sigma
  + \partial_{\mu}\bfpi\! \cdot\! \partial^{\mu}\bfpi)
  - \frac{1}{4} \lambda (\sigma^{2}+\bfpi\! \cdot \!
  \bfpi- \sigma_{0}^{2})^2 $} }

\newcommand{\EM}{\sqrt{k^2+m^{2}_{M}}}
\newcommand{\E}{\sqrt{k^2+m^{\star 2}}}
\newcommand{\EEE}{\bigl(k^2+m^{\star 2}\bigr)^{3/2}}
\newcommand{\EE}{\mbox{$ \sqrt{ k^{2}+(m-g_{\sigma}\sigma)^{2} } $}}
\newcommand{\ms}{\mbox{$ m-g_{\sigma}\sigma  $}}
\newcommand{\Ethree}{(k^2+m^{\star 2})^{3/2}}
\newcommand{\LLL}[1]{ {\cal L}_{#1}^{0} }
\newcommand{\gs}{ g_\sigma }
\newcommand{\gth}{ g_{3\sigma} }
\newcommand{\gfo}{ g_{4\sigma} }
\newcommand{\gv}{ g_\omega }
\newcommand{\gr}{ g_\rho }
\newcommand{\UU}{\mbox{$
 \frac{1}{3} b m (g_{\sigma} \sigma)^{3}
+ \frac{1}{4} c(g_{\sigma}\sigma)^{4}  $}}
\newcommand{\U}{\mbox{$
 \frac{1}{3} b m_{n} (g_{\sigma} \sigma)^{3}
+ \frac{1}{4} c(g_{\sigma}\sigma)^{4}  $}}
\newcommand{\Um}{\mbox{$
-\; \frac{1}{3} b m_{n} (g_{\sigma} \sigma)^{3}
- \frac{1}{4} c(g_{\sigma}\sigma)^{4}  $}}
\newcommand{\UUm}{\mbox{$
-\; \frac{1}{3} b m (g_{\sigma} \sigma)^{3}
- \frac{1}{4} c(g_{\sigma}\sigma)^{4}  $}}
\newcommand{\Upm}{\mbox{$ -\;b m_n (g_{\sigma} \sigma)^2
-c (g_{\sigma} \sigma)^3  $}}
\newcommand{\M}{\mbox{$ \sqrt{ k^{2}+(m_{B}-g_{\sigma B}\sigma)^{2} } $}}
\newcommand{\Mstar}[1]{\mbox{$ \sqrt{ k^{2}+\mstar{#1}^2 } $}}
\newcommand{\m}{\mbox{$ \sqrt{ k^{2}+m_{\lambda}^{2} }  $}}
\newcommand{\ke}[1]{\mbox{$ \frac{1}{2} m_{#1}^{2} #1^{2}  $}}
\newcommand{\keo}{\mbox{$\frac{1}{2} m_{\omega}^2 \omega_{0}^{2} $}}
\newcommand{\kerho}{\mbox{$\frac{1}{2} m_{\rho}^2 \rho_{03}^{2} $}}
\newcommand{\meff}{m_{B}-g_{\sigma B}\sigma}
\newcommand{\coup}[1]{\mbox{$ (g_{#1}/m_{#1})^{2} $}}
\newcommand{\lsigma}{\mbox{$\frac{1}{2}(\partial_{\mu} \sigma
\partial^{\mu} \sigma - m_{\sigma}^{2} \sigma^{2})$  }}
\newcommand{\lsigmaa}{\mbox{$\frac{1}{2}(\partial_{\mu} \sigma_{1}
\partial^{\mu} \sigma_{1} - m_{\sigma}^{2} \sigma_{1}^{2})$  }}
\newcommand{\lsigmab}{\mbox{$\frac{1}{2}(\partial_{\mu} \sigma_{2}
\partial^{\mu} \sigma_{2} - m_{\sigma}^{2} \sigma_{2}^{2})$  }}
\newcommand{\lsigmastar}{\mbox{$(\partial_{\mu} \sigma^\ast
\partial^{\mu} \sigma - m_{\sigma}^{2} \sigma^\ast \sigma)$  }}
\newcommand{\lomega}{\mbox{$ - \: \frac{1}{4} \omega_{\mu \nu} 
\omega^{\mu \nu} +\frac{1}{2} m_{\omega}^{2} \omega_{\mu} \omega^{\mu} $}}
\newcommand{\lrho}{\mbox{$ 
- \: \frac{1}{4}\bfrho_{\mu \nu}\! \cdot\! \bfrho^{\mu \nu} 
+ \frac{1}{2} m_{\rho}^{2}\bfrho_{\mu}\! \cdot\! \bfrho^{\mu} $}}
\newcommand{\omunu}{\mbox{$ \omega_{\mu \nu} = 
\partial_{\mu} \omega_{\nu}
- \partial_{\nu} \omega_{\mu} $}}
\newcommand{\nbaryon}[1] { \mbox{$ 
    \bigl(  \exp [(\epsilon_{B}(k) {#1})/T] +1  \bigr) ^{-1} $ } }
\newcommand{\nmeson}[2] {\mbox{$
    \bigl(  \exp [(\omega_{{#2}} (k) {#1})/T] -1 \bigr) ^{-1} $ } }
\newcommand{\D} {\mbox{$ {\cal D}(k,\omega) $} }
\newcommand{\Dm} {\mbox{$ {\cal D}^{-1}(k,\omega) $} }
\newcommand{\slpartial} {\mbox{$ \not\!\partial $} }
\newcommand{\slgamma} {\mbox{$  \not\!\gamma $} }
\newcommand{\sldel} {\mbox{$  \not\!\!\nabla $}}
\newcommand{\slp} {\mbox{$  \not\!\! p  $}}
\newcommand{\slA} {\mbox{$  \not\!\! A  $}}
\newcommand{\slK} {\mbox{$  \not\!\! K  $}}

\newcommand{\tit}
{Stangeness in Compact Stars and Signal of Deconfinement}
\newcommand{\autha} {Norman K. Glendenning}
\newcommand{\lbl}{\begin{flushright} LBL-40606\\[7ex] \end{flushright}}
\newcommand{\dateofdoc}{July 24, 1997}
\newcommand{\adra} 
{Nuclear Science Division \& 
Institute for Nuclear and Particle Astrophysics,
  Lawrence Berkeley Laboratory,
   MS: 70A-3307, Berkeley, California 94720}
\newcommand{\doe}
{This work was supported by the
Director, Office of Energy Research,
Office of High Energy
and Nuclear Physics,
Division of Nuclear Physics,
of the U.S. Department of Energy under Contract
DE-AC03-76SF00098.}

\newcommand{\ect}{A part of this work was done at the ECT*,
Villa Tambosi, Trento, Italy.}

\begin{titlepage}
\lbl
\begin{center}
\begin{Large}
\renewcommand{\thefootnote}{\fnsymbol{footnote}}
\setcounter{footnote}{1}
\tit {\footnote{\doe}}\\[5ex]
\end{Large}

\begin{large}
\autha\\[3ex]
\end{large}
\adra\\[3ex]
\dateofdoc \\[3ex]
\end{center}

\begin{figure}[htb]
\vspace{-1.2in}
\begin{center}
\leavevmode
\hspace{-1.2in}
\psfig{figure=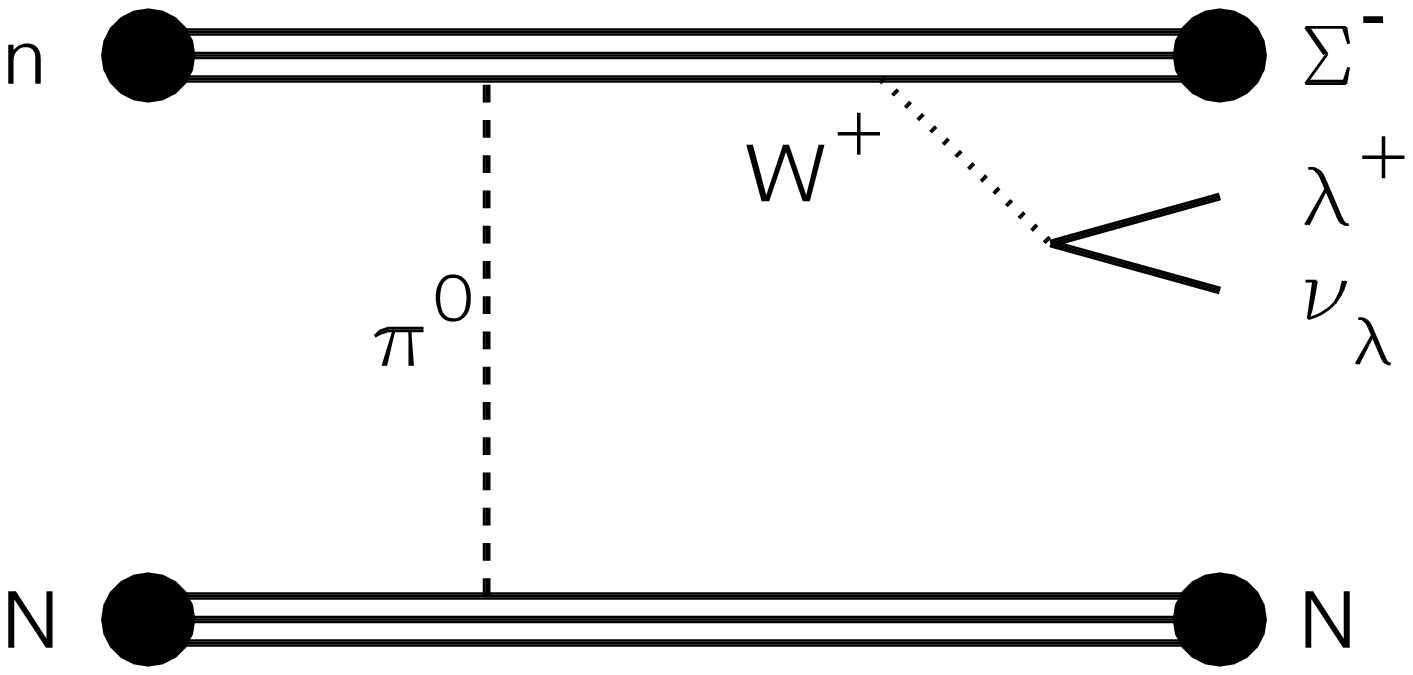,width=4in,height=2.66in}
\end{center}
\end{figure}


\begin{quote}
\begin{center}
 Invited Talk at the International Symposium on\\
 {\bf Strangeness In Quark Matter 1997},
 Thera (Santorini), Hellas\\
 April 14-18, 1997\\
 To be published in Journal of Physics G\\
 (Chair of the organizing committee: A. Panagiotou)
 \end{center}
 \end{quote}

\end{titlepage}

\clearpage

\clearpage
\begin{center}
\begin{Large}
\renewcommand{\thefootnote}{\fnsymbol{footnote}}
\setcounter{footnote}{1}
\tit \\[7ex]
\end{Large}
\renewcommand{\thefootnote}{\fnsymbol{footnote}}
\setcounter{footnote}{2}

\begin{large}
\autha \\[2ex]
\end{large}
\adra
\end{center}


\begin{quote}
Phase transitions in compact stars are discussed including hyperonization,
deconfinement and crystalline
phases. Reasons why kaon condensation is unlikely is reviewed.
Particular emphasis is placed on the evolution of internal structure
with spin down of pulsars. A signature of a first order phase
transition in the timing structure of pulsars which is strong and easy
to measure, is identified.

\end{quote}

\section{Introduction}

During their study of supernovae, Baade and Zwicky suggested in 1934
that the enormous energy that is released so suddenly as to make
even distant objects visible in daylight and for weeks thereafter
must originate
in the transition of the core of an ordinary
star to a neutron star---a star    consisting of closely packed neutrons.

By an elementary calculation of the type one
learns in potential theory, one can estimate the gravitational energy
of a star of closely packed neutrons and find that a nucleon in a 
neutron star is bound 10 times more strongly by the weakest force,
gravity, than a nucleon is bound in a nucleus by the strong
force \cite{book}. The implication of this statement is immediate. A star,
when it is exploded by the binding energy of the newly formed
neutron star releases in a few seconds about 10 times as much energy
as the luminous star emitted in its lifetime of a few million years.

As was expected, and confirmed in observations
made on SN1987A, about 99 \% of the binding energy
is carried off by neutrinos emitted over about 10 seconds from the hot
protoneutron star. Of the remaining 1 \% almost all is carried by the
kinetic energy of the
stellar remnant as it is hurled into space at a velocity of 12,000 km/s. 
The remaining fraction of the 1 \% is in visible light.

These are wondrous thoughts and they inform us
that the star that is formed at the death of a luminous
star is made of matter under extreme conditions. 
In the year 1047
the  Chinese court
astronomer 
of the Sung Dynasty ``observed the apparition of a guest star ....
its color an iridescent yellow...''  This was a part of
the 1/10 \% of energy  that appeared as visible light.
It was the Crab supernova
visible today,  much expanded and  
{\sl accelerating}. The acceleration is powered
by the rotational energy
of the Crab pulsar, the neutron star formed at the end of the 
\begin{figure}[htb]
\vspace{-.4in}
\begin{center}
\leavevmode
\psfig{figure=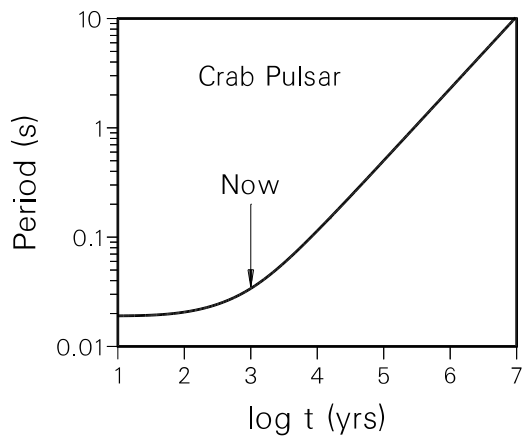,width=3.3in,height=2.76in}
\parbox[t]{4.6 in} { \caption { \label{crab25} Evolution of the Crab pulsar's
period according to the measured braking index.
}}
\end{center}
\end{figure}
10 million year life of the presupernova star.
From observations on the nebula,
it can be inferred that the  power input by the pulsar  is
$10^{44}$ MeV/s.
From this one compute the  moment of inertia
and then the  rotational energy
as  $10^{55}$ MeV. The Crab pulsar, rotating now 30 times a second,
will  be rotating in a million years once a second
(see Fig.\ \ref{crab25}). And still have
some rotational energy left. We will exploit this slow
rate of spin-down  to develop a diagnostic for the deconfinement
phase transition.

\begin{figure}[tbh]
\vspace{-.5in}
\begin{center}
\leavevmode
\centerline{ \hbox{
\psfig{figure=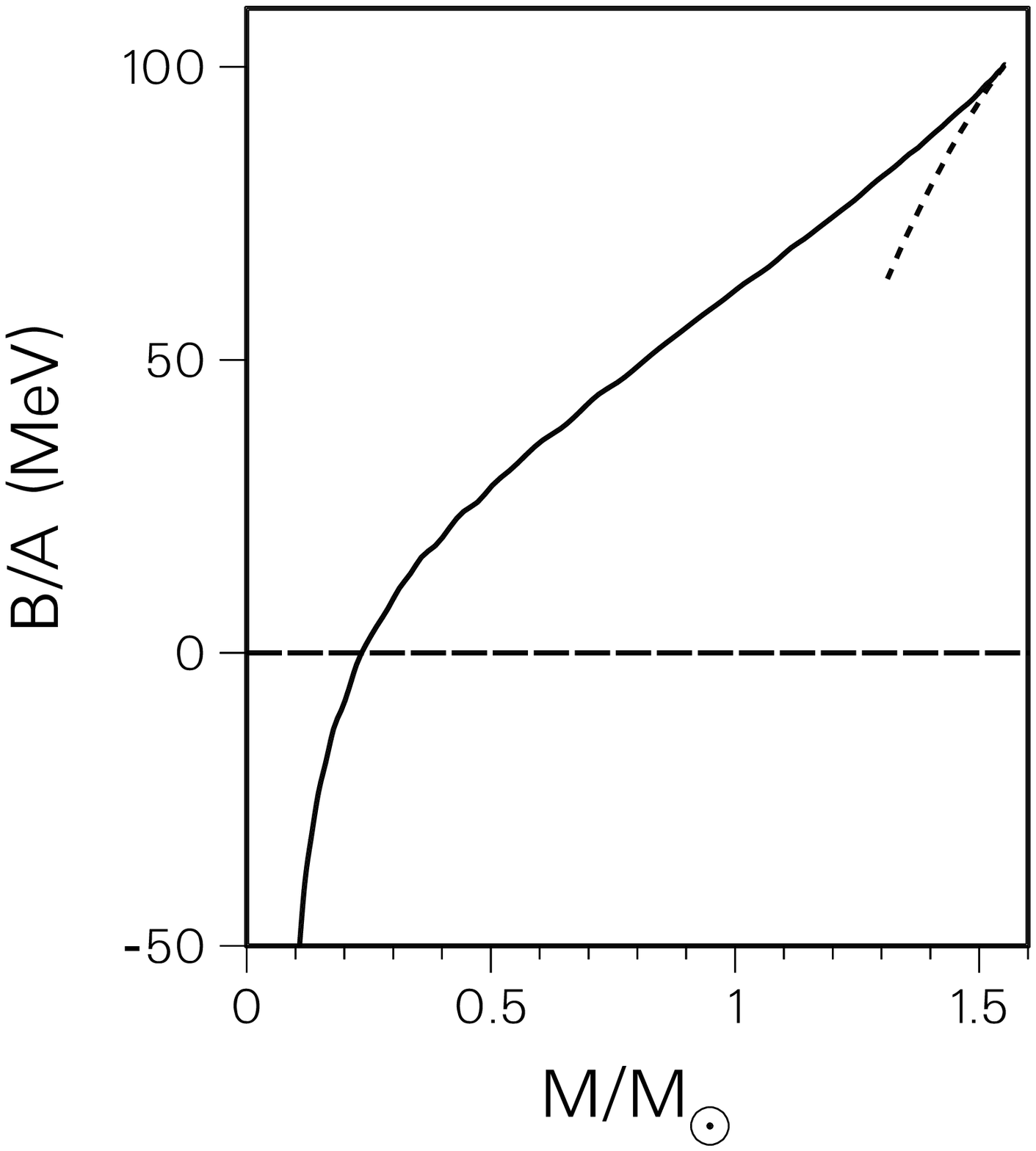,width=2.5in,height=3in}
\hspace{.6in}
\psfig{figure=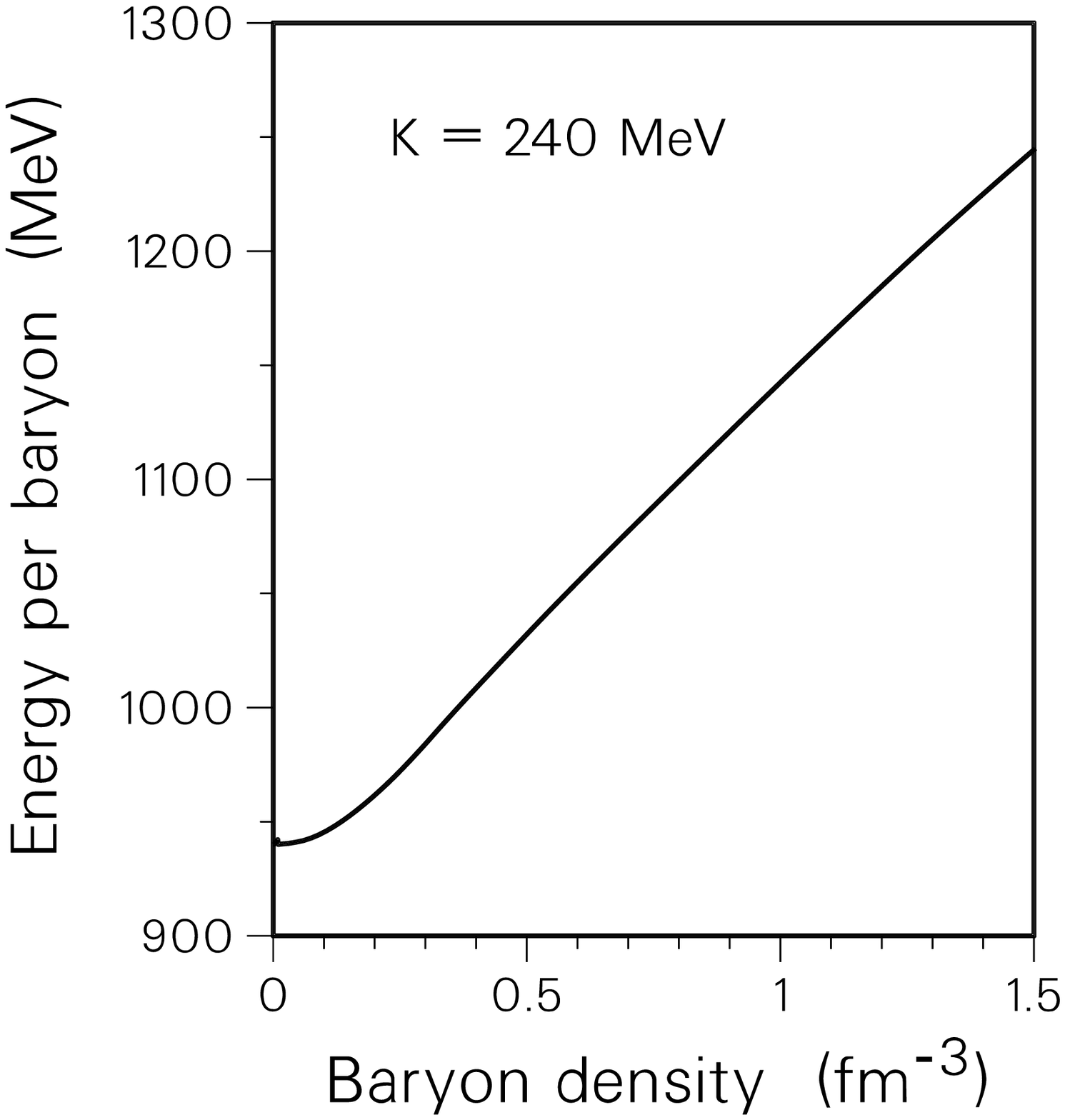,width=2.5in,height=3in}
}}
\begin{flushright}
\parbox[t]{2.7in} { \caption { \label{bvsm} Binding energy per nucleon.
Effectively this is the gravitational energy less the average
compression energy \protect\cite{book} \courtesy.
}} \ \hspace{.4in} \
\parbox[t]{2.7in} { \caption { \label{bam} Compression energy.
Typically the compression energy of matter at the stellar
center is $\sim 250$ MeV. The average compression
is typically $\sim 100$ MeV.
}}
\end{flushright}
\end{center}
\end{figure}

What is implied by the above observations  about the internal 
structure and composition of neutron stars? I have already
noted that gravitational binding is 10 times nuclear binding
(see Fig.\ \ref{bvsm}). 
The average density of
neutron stars is several times nuclear density. This we can infer by
balancing gravity against the centrifugal force at the surface of
a millisecond pulsar. So nucleons reside within the repulsive interaction
of their neighbors; gravity binds the star.
 The nuclear force resists
gravity but fails; neutron stars exist. In crushing nucleons to
densities where nuclear repulsion is dominant
(see Fig.\ \ref{bam}), gravity brings the
Pauli principle fully into play in distributing baryon charge over
as many species such as hyperons and quarks as is energetically
favorable. The name neutron star therefore has to be understood
as generic.  

Stars rotate. We know this from the doppler broadening of the spectral 
lines of luminous stars. We infer it also from the periodic radiation
from pulsars.
If the periodicity were caused by vibration, the
amplitude would decrease with time and the frequency remain constant. 
What is seen is the converse. The conclusion that what we see from pulsars
is the beamed radiation of a rotating source makes eminent physical sense.
The amplitude of
vibrations would soon be damped by viscosity.
Rotation is damped very weakly by coupling to the
electromagnetic field and not at all by viscosity. 
This accords with the observed small rates  of change of pulsar
periods.  A pulsar could be seen for only a short time
in the first case; for millions of years in the second.
If the pulsar signal were caused by vibration
the observed pulsar population would imply the existence of an enormous
number of silent neutron stars and their number
would be irreconcilable with supernova 
rates.

Stars have magnetic fields. This we know from the Zeeman effect.
Flux conservation implies
a very high field for the collapsed remnant.
So the periodic signal from a pulsar
is caused by the rotation
of a magnetized neutron star. Like the magnetic field, the 
angular velocity  is scaled up in the collapse, 
sometimes to tens of times  per second. They slow
down very slowly, taking 10 million years to do so. During the stage of
rapid rotation they are centrifugally flattened. As they spin down
due to the radiation their central density rises and they become 
more spherical. Through
the Pauli principle, the increase of density with time induces
continuous changes in internal composition. 
Like us, their faces change continuously. Like us they also may become
more interesting with age.

I will tell you about some of the many ways in which the Pauli principle
can act to create the many faces of neutron stars and the way in which
spin-down of a pulsar may be used to detect some of the faces.

\subsection{Charge Neutrality}

The first thing I should say about stars---any star---is that to 
a very high degree they are charge neutral. This follows at once
if we examine the balance between gravitational attraction on a charged
particle at the surface of a star and the  Coulomb repulsion
acting on it. We find for a proton that 
$$ \frac{ Z_{{\rm net}} } {A } < \biggl( \frac{m}{e} \biggr)^2
\sim 10^{-36}\,.$$
This does not mean that no charged particles above this small number 
can exist in the star. It only means that the number of positive and
negatively charged particles must be nearly equal. Nevertheless it
is a very stringent condition on the nature of stellar matter
as we  shall see.
Charge neutrality of a star
(more precisely, its isospin asymmetry)
also gives rise to a most amazing phenomenon---the
formation of a Coulomb lattice in the region of a star  occupied by
two phases in equilibrium---the so-called mixed phase of a first order
phase transition.

\subsection{Degenerate Matter}

Compact stars cool very rapidly after birth, falling from temperatures
of several tens of
MeV to an MeV or less in a few seconds as the neutrinos
carry the bulk of the  binding energy off. Thereafter they cool more slowly, 
but they are cold on the nuclear scale in any observational time-frame.
Therefore compact stars are degenerate Fermi systems.  The smearing at the
Fermi surface due to temperature is extremely small compared
to the Fermi energy. The Fermions 
fill all lowest momentum states up to 
the Fermi level. These momentum states are not the momentum states of
free particles because of the interactions. But for purposes of discussion
we may think of them as free momentum states of energy
and density
$$ E_p = \sqrt{m^2 + p^2},~~~~~~~\rho=\frac{1}{(2\pi)^3} \int_{0}^{p_F}
p^2 d^3 p \,.$$

\section{Hyperon Stars}

Originally,  compact stars were thought of as 
 closely packed neutrons.
Even to-day some researchers, especially those who seek to solve the
Brueckner-Bethe many-body theory with so-called realistic forces
treat neutron stars as purely neutron. However this is very unrealistic.
First we know from the valley of beta stability, that nuclear matter
prefers to be isospin symmetric. Gravity does not fully permit this as noted
above since excess charge would simply be blown away.
There is of course another principle, and that is that a fully developed
star should be in the ground state of cold matter. So the matter of a compact
star will therefore be in the lowest energy state---at each
density---consistent with charge neutrality. 

What does the above conclusion mean for the composition of neutron star
matter in first approximation? It means that a compact star cannot
be purely made of neutrons. Such a configuration would be neutral but 
not in the lowest energy state. Some neutrons at the top of the Fermi
sea would have
energy sufficient to inverse beta decay to a proton, electron and neutrino.
The neutrino, having no mass or very small mass, will have the escape 
velocity when it diffuses to the surface of the star. 
Lepton number is therefore not conserved and the star's energy is lowered
by their loss. 

How many neutrons will decay and to what? It depends on the density
so that the composition of neutron star matter is a continuously changing
function of density. Let us see how this works. At low density, neutrons
will decay to protons and electrons until the energy of the system cannot
be further lowered.  This situation is referred to as beta equilibrium
and is attained when the Fermi energies satisfy the relation
$$ \mu_n=\mu_p+\mu_e \,.  $$
This equation states that the momentum eigenstates are filled in such a way
that no energy can be gained or lost by a reaction in either direction.

\begin{figure}[htb]
\vspace{-1.2in}
\begin{center}
\leavevmode
\hspace{-1.2in}
\psfig{figure=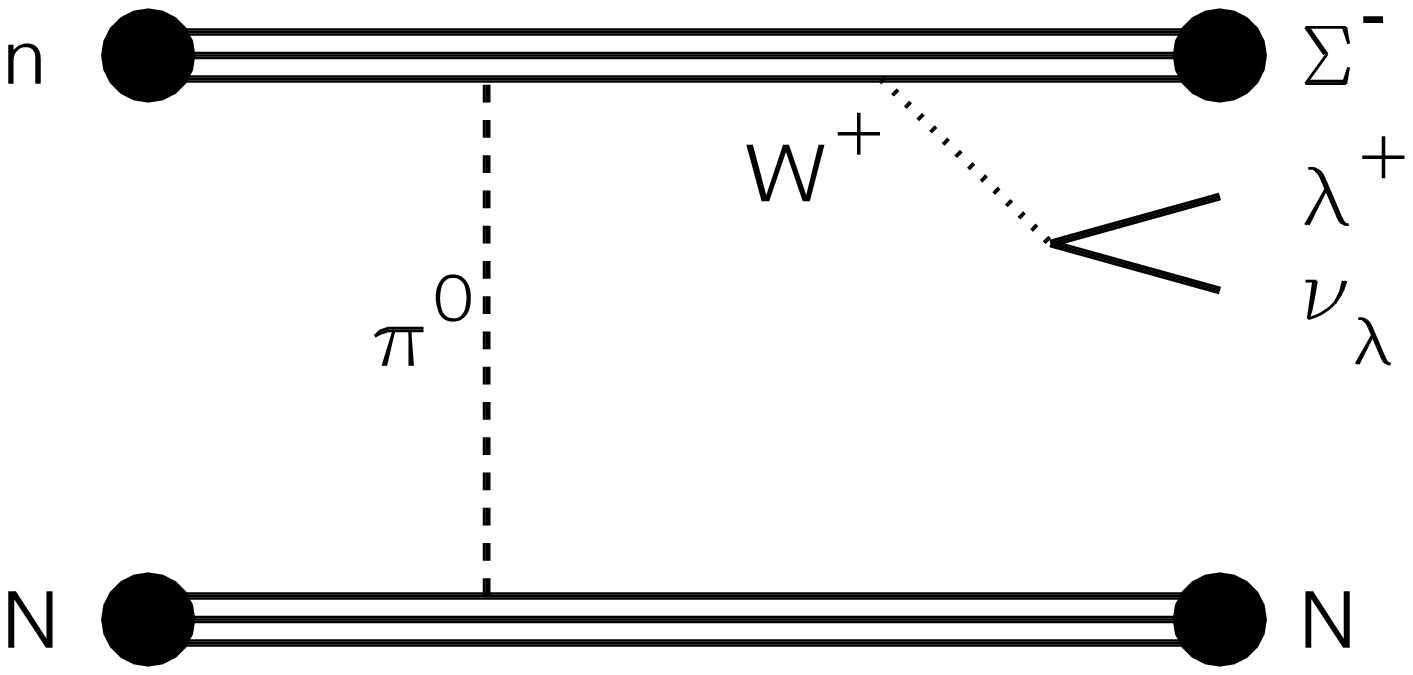,width=4in,height=2.66in}
\vspace{-.2in}
\parbox[t]{4.6 in} { \caption { \label{weak} Weak interactions such as
the one depicted are of little importance in vacuum because of their
strong suppression but carry a neutron star
to its final equilibrium state in the confined phase when there is
insufficient energy to produce an associated kaon.
}}
\end{center}
\vspace{.2in}
\end{figure}

The Fermi momentum 
of the neutron will increase with increasing density
toward the interior of a star ($k_{{\rm F}}\propto \rho^{1/3}$). Eventually,
the neutron Fermi energy will be so high that the neutron can decay also to
other baryon species besides the proton. It can decay for example to
the $\Lambda$ or $\Sigma^-$, etc.
You may object that strangeness is changed by such decays. However it
is only the strong interaction that conserves strangeness. The weak
interaction violates it. For example,
$$ \mu^- + K^+ \longrightarrow \mu^- +\mu^+ + \nu  \longrightarrow
2\gamma +\nu \,.$$
The gammas and neutrinos diffuse out of the star lowering its energy.
Reactions become irreversible. Net strangeness is built up
through  reactions like the strong followed by weak interaction 
$$ N+N +\mu^- \longrightarrow N+\Sigma^- + K^+ \mu^-
\longrightarrow N+\Sigma^- + 2\gamma +\nu
\,, $$
or by the weak interaction (see Fig.\ \ref{weak})
$$ N+N+\mu^- \longrightarrow N+\Sigma^- + 2\gamma +\nu $$
The strong  reaction followed by the
weak can occur when the protoneutron star is far
from equilibrium, and the direct weak conversion
of a nucleon to hyperon will carry the system to final 
equilibrium when there is insufficient energy to produce the 
intermediate state containing the kaon.
You see at each point the Pauli principle
drives neutral matter from a predominance of neutrons at low density
to an ever richer mixture of baryon species, especially hyperons
at high density where the increasing
Fermi energy   exceeds the thresholds
for conversion of neutrons to other more massive baryon  species.
 The conclusion is robust since it rests principally
on the exclusion principle.

\section{Neutron Star Matter}

I discussed hyperonization and kaon condensation
 in 1985  and found that hyperonization has a strong 
effect on limiting the maximum mass that a neutron star can have, and that
kaon condensation was most likely prevented by hyperonization
\cite{glen85:b}.
I will recall the reasons for this in a latter section.

The covariant Lagrangian that I generalized to include hyperons
(ie. the entire baryon octet) from earlier 
work of Johnson and Teller \cite{teller55:a},  Duerr
\cite{duerr56:a},
in the mid 1950's  and
and Walecka \cite{walecka74:a} in 1974 is
\begin{eqnarray}
{\cal L} & = &
\sum_{B} \overline{\psi}_{B} (i\gamma_{\mu} \partial^{\mu} - m_{B}
+g_{\sigma B} \sigma  - g_{\omega B} \gamma_{\mu} \omega^{\mu}
- \fraca g_{\rho B} \gamma_{\mu} \bftau \cdot \bfrho^{\mu} )
{\psi}_{B} \nonumber  \\
&   &+\:  \lsigma
  \lomega  \nonumber\\[2ex]
  &   &  \lrho
    \:\Um  \nonumber\\[2ex]
    &   &  + \sum_{e^{-},\mu^{-}}
    \overline{\psi}_{\lambda} \bigl(i\gamma_{\mu}
      \partial^{\mu} - m_{\lambda} \bigr) \psi_{\lambda}\,. \nonumber
	\label{lagrangian}
	\end{eqnarray}
I certainly do not want to discuss the theory
in any detail. First of
all I think it should be emphasized that we have very little knowledge
of matter at high density and it seems to me that there
is a limit to how much effort can be 
invested meaningfully in elaborating theories of dense matter.  
I choose instead a simple theory guided by  the 
 following principles:
\begin{enumerate}
\item Compact stars are relativistic and should be described in the
framework of General Relativity.
\item Matter is causal and should be described by a relativistically 
covariant theory.
\item Stellar matter at any density obeys the condition of
microscopic stability (le Chatelier's principle).
\item At high  density (or momentum transfer) quarks are asymptotically
free.
\item Theory of dense matter should be constrained by what is know of
nuclear matter at saturation (binding, saturation density, compressibility,
symmetry energy, nucleon effective mass).
\item Since nuclei provide no information on the hyperon sector,
constraints on hyperon couplings should be obtained from data
on hypernuclei and the binding of the $\Lambda$ in nuclear matter.
\end{enumerate}
I believe that what is allowed by the laws of nature is quite possibly
realized somewhere in the universe. It is in that spirit that I 
study compact stars and their many faces---hyperon stars,
hybrid stars, strange matter stars...., constrained 
by the above principles and data.

\begin{figure}[tbh]
\vspace{-.5in}
\begin{center}
\leavevmode
\centerline{ \hbox{
\psfig{figure=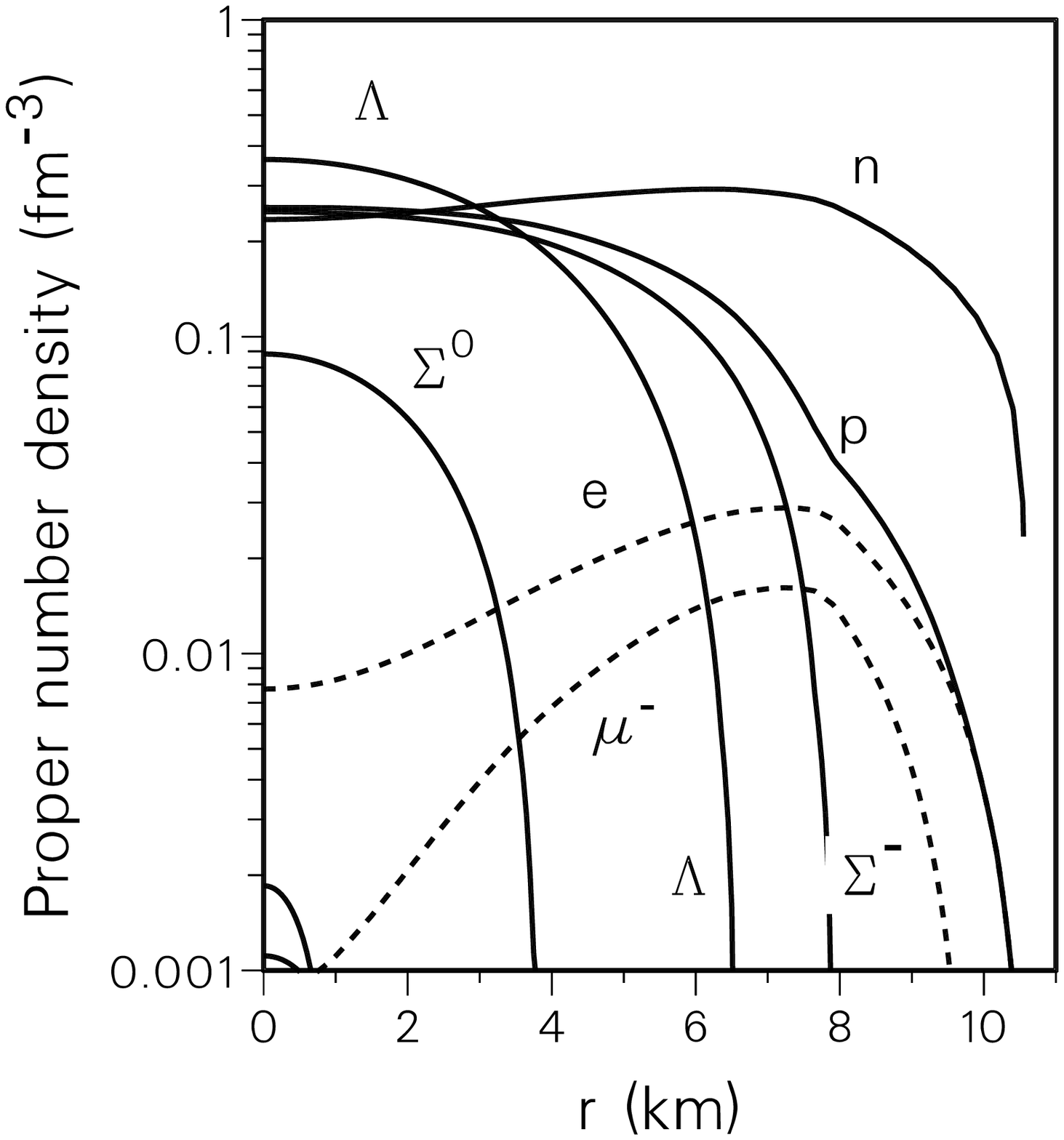,width=2.5in,height=3in}
\hspace{.6in}
\psfig{figure=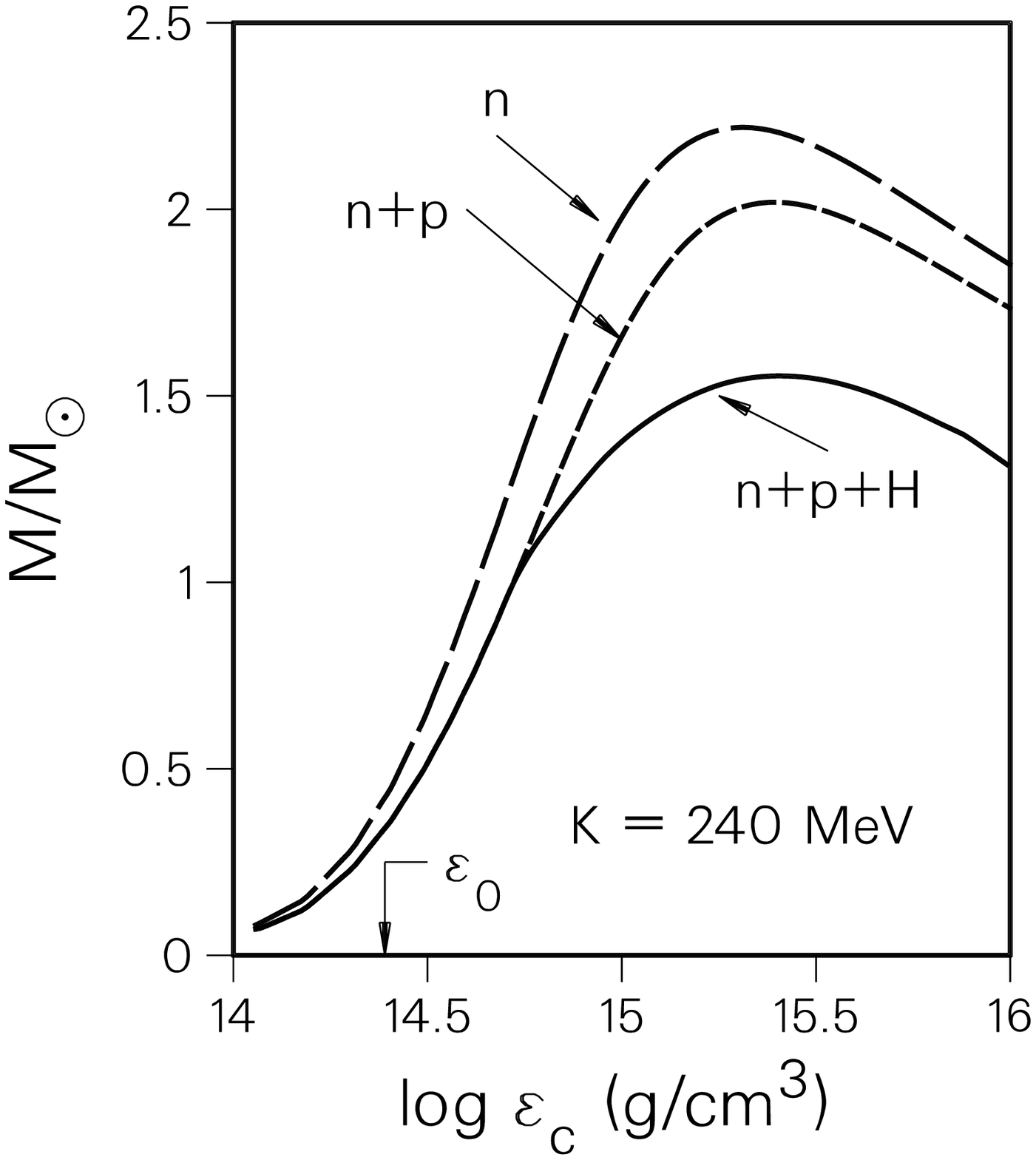,width=2.5in,height=3in}
}}
\begin{flushright}
\parbox[t]{2.7in} { \caption { \label{comp_k240} Baryon populations in 
a hyperon star \protect\cite{glen91:c}.
}} \ \hspace{.4in} \
\parbox[t]{2.7in} { \caption { \label{mass_3} Stellar sequences
labelled by the completeness of the description of hadronic matter
\protect\cite{book} \courtesy.
}}
\end{flushright}
\end{center}
\end{figure}

\section{Composition of Hyperon Stars}

The result of incorporating  hyperons
into a description of neutron stars is illustrated
in Fig.\ \ref{comp_k240}. Neutrons remain the
majority at 60\% but protons and strange baryons each make up about
20\% in the maximum mass star
\cite{glen91:c}. 

Near the edge of the star, neutrons
are the dominant species as can be seen
from Fig.\ \ref{comp_k240}, but at a depth of several kilometers,
the $\Sigma^-$ is populated, and at all higher densities,
the electron population is suppressed. This is because baryon number in the
star is conserved but not lepton number (neutrino diffusion).
Consequently, with rising density, charge neutrality can
be achieved among charged particles carrying baryon charge with little
need for electrons. Therefore the electron population
saturates and falls with increasing density. (See Fig.\ 2 in Ref.\
\cite{glen85:b} and the discussion  of
kaon condensation of section III.)
Electron saturation has an important consequence for kaons, as was 
explained already in 1985 \cite{glen85:b}, kaons would  be
 energetically 
favorable substitutes for electrons {\sl if} the electron chemical potential
(Fermi energy) were to
exceed the kaon mass in the medium. In such an event,
kaons, being bosons, could all occupy the lowest  
energy state (having zero momentum)
whereas electrons, being Fermions cost the Fermi energy.
However, when
charge neutrality can be achieved  among the   baryons, whose
number is fixed at the birth of the star, not even
the cost of  a reduced
kaon effective mass  has to be paid.  Deconfinement has the same effect on the
electron chemical potential as charged baryons and for the same reason.

Three stellar sequences are shown
in Fig.\ \ref{mass_3}, corresponding to different
degrees of completeness  of the calculation with respect to beta
equilibrium. Some of the
possible consequences of 
this revised   picture of a ``neutron'' star are:
\begin{enumerate}
\item The maximum mass that the Fermi pressure and the repulsion of
the nuclear force can sustain against gravity will be reduced, because
 the  transformation of nucleons at the top of their
 Fermi sea to hyperons  reduces the pressure, that is to say,
 softens
the \eosp.
This effect, first discussed in Ref.\ \cite{glen85:b}, was refined
in Ref.\ \cite{glen91:c}. The estimated
limiting neutron star mass
lies in the range $1.5 {\rm~to~}2 \msun$ and its radius lies in the
range $\sim 10.5 {\rm~to~} 12$ km. However, as stated above, theory
is only constrained by general principles at high density, and the quoted
numbers, as with any others, have to be understood in that context.
\item It is usually thought that at the end of its life a star
will either collapse entirely to from a black hole equal to its mass, say
10 to 50 $\msun$ {\sl or} a small fraction of the binding energy of the
core will be transmitted by neutrinos
to the infalling star and explode it leaving behind a neutron
star. There is another 
possibility suggested by the Fig.\
\ref{mass_3}. If the baryon number of the collapsing core produces a
compact star 
whose mass lies above
the limiting mass of a hyperon star but below that of a neutron star
(which we may use to approximate a protoneutron
star before it has cooled and deleptonized),
the core will hover in a hot bloated state for a few seconds while
many of the neutrinos escape, blowing off the rest of the star
as they do so, and then
the core will collapse to a low-mass black hole of 1.5 to 2 $\msun$
\cite{janka95:a,glen94:d,prakash95:b}.
\item We expect therefore two populations of black holes (1) massive ones
of tens of solar masses that swallow up the entire progenitor star,
neutrinos included (no supernova) (2) Low-mass black holes  
accompanied by a neutrino burst and a SN that  distributes  enriched material
to the cosmos.
\item Transport properties like electrical conductivity, superfluity,
and cooling of the star will be effected by the hyperon populations. 
\end{enumerate}

\section{Hybrid Stars}

There is
another possibility, not exclusive of the hyperon nature of dense
neutral matter. If  densities are attained in the centers
of compact stars that are high enough to  crush  nucleons
into their quark constituents then quark matter, 
which existed in the early universe
as very hot matter, may exist in a cold state in the cores of
hyperon stars. I call these hybrid stars \cite{glen91:d,glen91:a}.

The possibility of quark matter interiors was discussed already in
1976 by Baym and others, and was investigated right up to the present.
However I discovered in 1990 that a very important aspect of phase
transitions was overlooked in all of the early work, including my own.

The consequences that follow from a correct treatment of the phase transition
are quite astonishing. We find that a remarkably intricate crystalline
lattice of nuclear and quark matter is formed, that changes in size and
form in various regions of the star according to the ambient pressure.

I will not go into a full description of the theorems that I proved.
But I will briefly state them and physically motivate them.
I recall to you first what is very familiar. You know that when ice
begins to melt, as the day warms, the temperature of the mixture of 
ice and water remains constant until the ice is completely transformed.
Then the water can warm.

What may not be so familiar is that not only does the
temperature of water and ice remain constant but so do all
their other  properties.
The density of ice and the density of water remain constant, for example,
until the ice is completely melted. 
If the same experiment were performed at constant temperature but varying
pressure, it would be found that the pressure remained constant until
the transformation was complete. This is all very familiar; we know it from 
our courses in thermodynamics and statistical physics. What is not familiar
is that the above description of a first order phase transition is
a very special case. It pertains to substances that have a single independent
component (or conserved charge)---in the above example, H$^2$O.

The general case of a first order phase transition in a (complex)
substance of more
than one independent component is very different. I proved the following
theorems {\cite{glen91:a}:
\begin{enumerate}
\item 
{\sl Any} first order phase transition in a complex substance
is characterized by
a variation of the common pressure, and the variation of every property
of the two phases in equilibrium as the proportion of the phases changes.
This is the exact antithesis of the phase transition in the 
single-component substance that I just described.
\item If one of the independent components is electric charge, then
the two phases will in general have opposite charges and the 
energy of the mixed phase will be minimized when the rare phase arranges
itself on a lattice immersed in the dominant phase.
\item Because of theorem 1, the lattice will vary in form and size with
proportion of the phases.
\end{enumerate}

Words cannot describe wonders as well as images. Let me give you an image.
If water were of such a nature as described, ie. had two instead of
one independent component and one of them was electric charge, then
a lake would not freeze over starting with a sheet of ice on top, but
ice spheres would form throughout the volume of the lake, of slightly
different size and spacing at top and bottom, because of the different
pressure. That is a valid image for neutron star matter when it is in the
pressure gradient of the star and it is
of such a density that quark matter and nuclear matter are in equilibrium.
\begin{figure}[tbh]
\begin{center}
\leavevmode
\centerline{ \hbox{
\psfig{figure=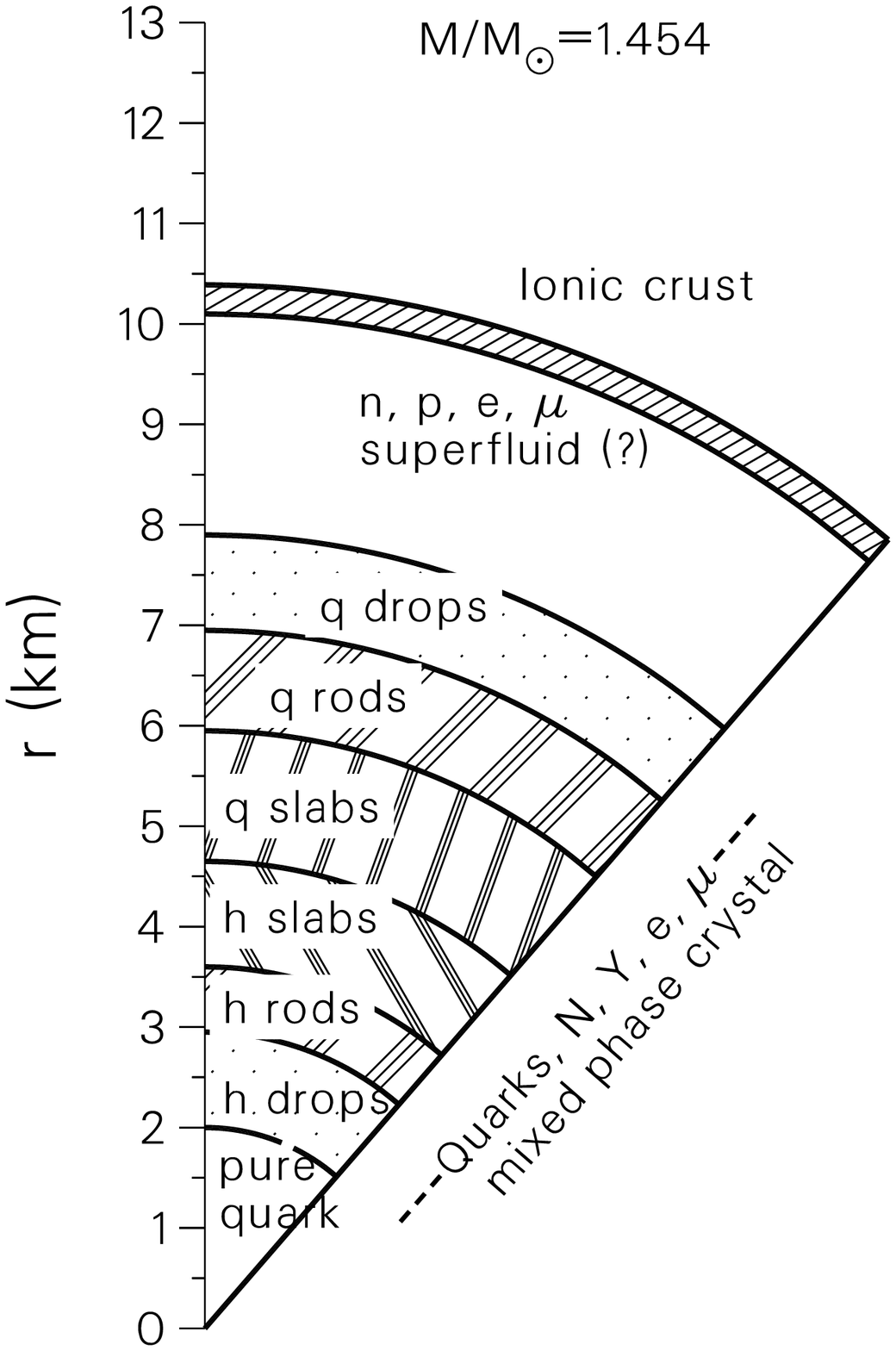,width=2.5in,height=3.in}
\hspace{.5in}
\psfig{figure=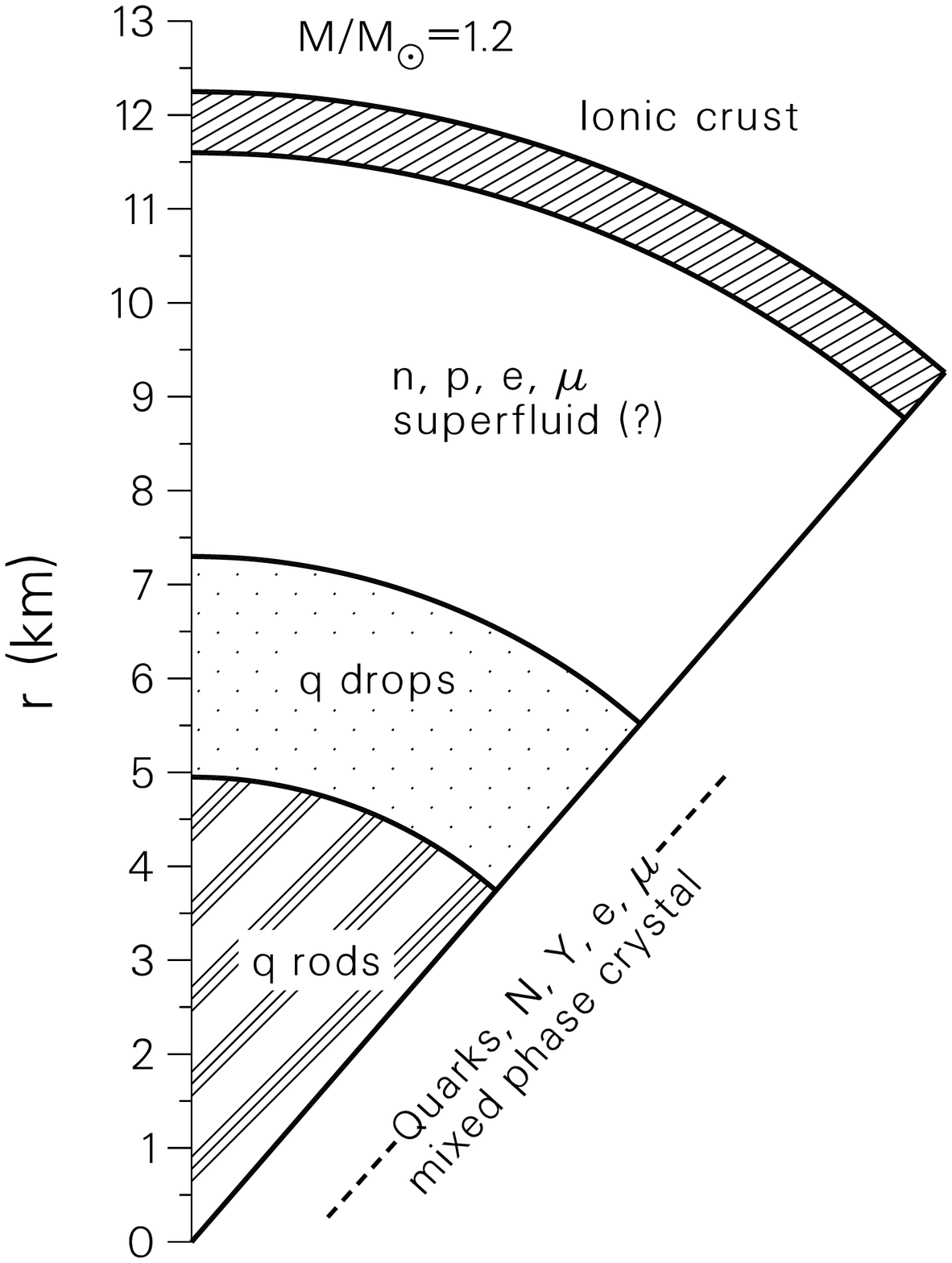,width=2.5in,height=3.in}
}}
\vspace{.5in}
\begin{flushright}
\parbox[t]{2.7in} { \caption { \label{pie} Showing the quark gas
core, surrounding crystalline region, hyperon liquid and thin
nuclear crust. Geometric phases are denoted a q(uark) drops, etc.
\protect\cite{book} \courtesy.
}} \ \hspace{.18in} \
\parbox[t]{2.7in} { \caption { \label{pie2} For a slightly less massive
star than depicted in Fig.\ \protect\ref{pie}  the interior structure
is vastly different. The Coulomb lattice extends to the center, but only
several geometrical phases are present \protect\cite{book} \courtesy.
}}
\end{flushright}
\end{center}
\end{figure}
If I may say so, I find this a remarkable revelation. I am astonished 
that so many researchers before me missed it. That in fact  it took me 
five years to arrive at this understanding which now seems so clear and
obvious.

I have proven the above theorems elsewhere, though I have not always
laid them out as theorems \cite{glen91:a,glen95:c,glen94:e,glen95:e}.
Here I want to provide only the physical understanding, which is just as
good as a mathematical proof, and usually more illuminating.
Nuclei become unbound around $A\sim 250$, and no other bound
configuration of nucleons exists until $A$ is so large that gravity
binds them. 
But neutron star matter  must be neutral
to be bound by gravity.
So while charge neutral nuclear matter
is highly isospin asymmetric and unfavored by
the strong force,
 such a configuration
 for a large number of baryons
is nonetheless
favored by the balance of forces since  gravity dominates.

 The symmetry driving forces in nuclear matter
 consist of the nuclear force itself (say the coupling of the rho
 meson to the nucleon isospin current) which favors symmetry,
 and a Fermi energy contribution which favors distribution of baryon charge
 equally over nucleon and proton (and at higher density over as
 many species as are energetically available). In quark matter, 
 which we think of as approximately free, only the Fermi contribution
 exists and  even
 at moderate density there are three Fermi seas available,
 those of the light quarks. 
 Consequently, when the density is raised to the point that some
 neutron star matter is converted to quark matter, the repulsive
 isospin restoring force in the nuclear matter can be relieved 
 by exchanging charge and possibly strangeness (mediated by the intermediate
 vector bosons) between
 regions of neutron star matter and those of quark matter in equilibrium
 with it. The degree to which the isospin repulsion can be reduced
 obviously depends on the proportion of matter in each phase in equilibrium
 and is restricted by the conservation laws.
   Therefore the energy is not a linear function of the proportion of
   phases in equilibrium  but has the
   form
   $$ E=(1-\chi)E_H(\chi) + \chi E_Q(\chi)$$
   where $\chi=V_Q/V$.
   Consequently the pressure is not a constant
   in the mixed phase (recall $P=-\partial E/\partial V$).
   Fig.\ \ref{chiq_k300b180} shows how the charge on the
   two phases changes continuously as a function of the proportion $\chi$
   of quark matter in the coexistence phase. Likewise all other properties
   vary. This schematic proof can be made rigorous \cite{glen91:a}.

   The impact of the above observation on neutron star structure is
   major. In the early work,  the mixed phase which had constant
   pressure independent of the proportion, was absent from the 
   monotonically varying pressure environment of stars and a large density
   discontinuity occured at the pressure corresponding to the mixed phase.
   These  artificial
   aspects were consequences of treating the star as a purely
   neutron star, which is beta-unstable, or imposing local charge
   neutrality on the solution of theory for a beta-stable star. As
   proven in Ref.\ \cite{glen91:a}, the conditions
   of local neutrality in the mixed phase are incompatible with Gibbs
   conditions for equilibrium. Conservation laws are global and in non-uniform
   systems (like the mixed phase) must be imposed on the solution of ones
   theory only
   in a global sense.
 Each phase in equilibrium
 can be charged; only the total change must vanish. 
 The early work on quark matter in neutron stars
 therefore missed the fascinating formation of spatial structure
 in the mixed phase that results from the competition of the
 Coulomb interaction and the surface interface energy.

\begin{figure}[tbh]
\vspace{-.5in}
\begin{center}
\leavevmode
\centerline{ \hbox{
\psfig{figure=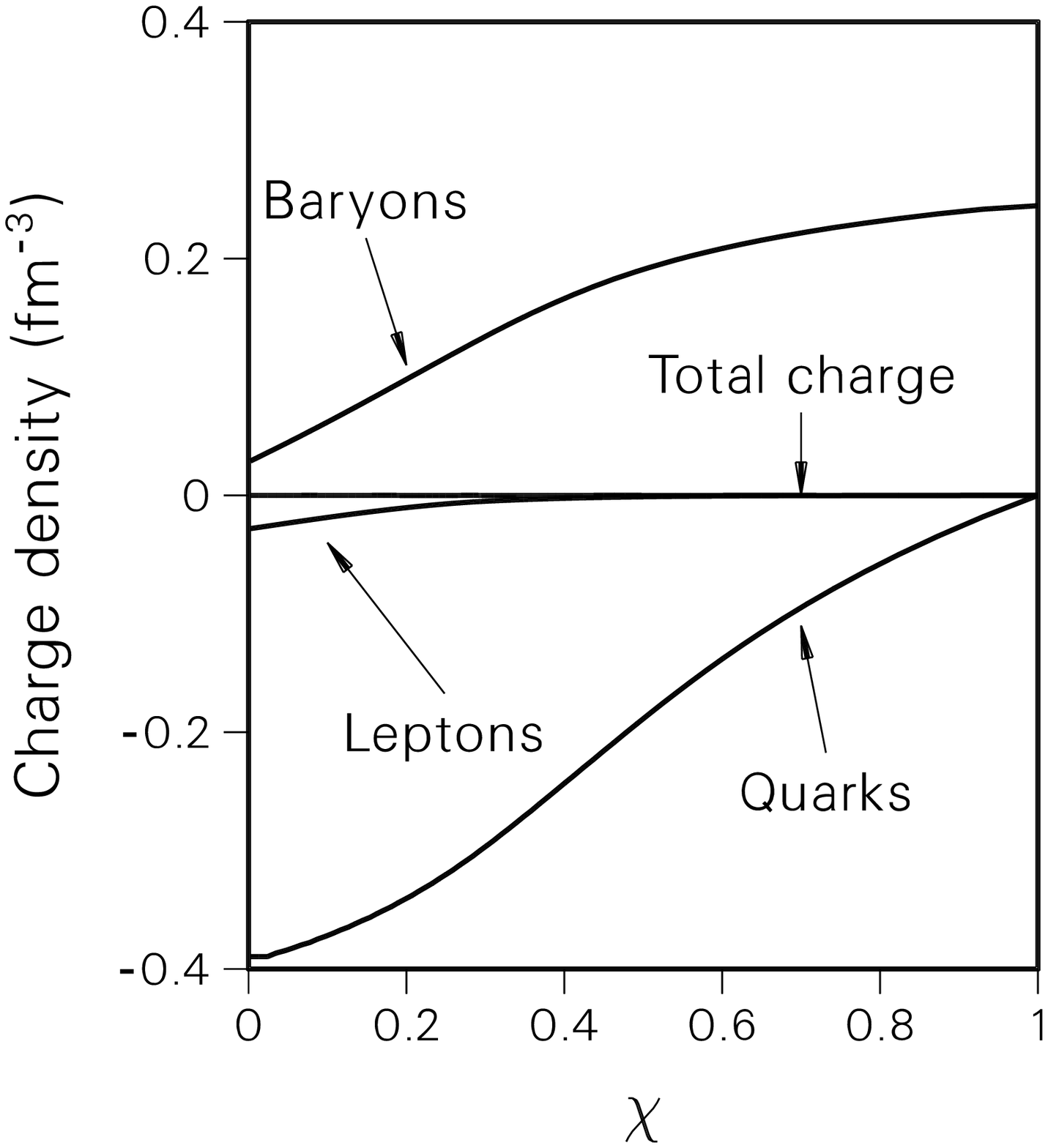,width=2.5in,height=3in}
\hspace{.5in}
\psfig{figure=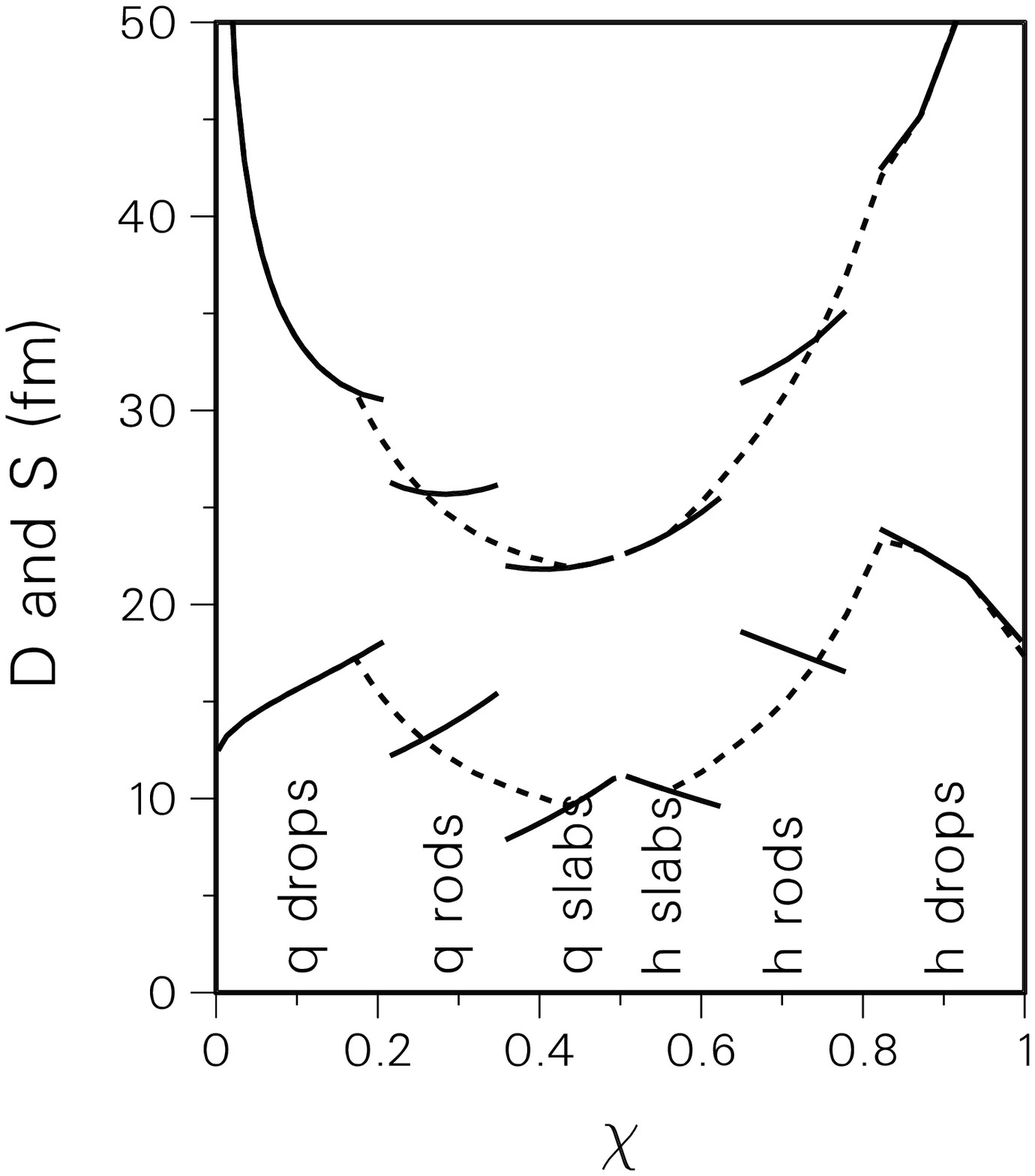,width=2.5in,height=3in}
}}
\begin{flushright}
\parbox[t]{2.7in} { \caption { \label{chiq_k300b180} Charge density
on regions of both phases in equilibrium and the uniform electron
density as a function of proportion of
quark phase \protect\cite{book} \courtesy.
}} \ \hspace{.18in} \
\parbox[t]{2.7in} { \caption { \label{crystal_k300b180} Diameter (D)
and spacing (S) of geometrical phases as a function of proportion of 
quark phase.
}}
\end{flushright}
\end{center}
\end{figure}

The key concepts are encapsulated in the words ``degree(s) of freedom''
and ``driving force(s)''.  The degree of freedom here is that of exchanging
charge to achieve an energetically more favorable concentration. 
The driving force is the isospin restoring force of nuclear matter,
composed of contributions from the Fermi energy and explicitly through
the interaction of baryons with the rho meson.

\section{Crystals in Stars}

Now turn to the proof of the last two theorems. The symmetry restoring force
causes the rearrangement of charge
so that nuclear matter has positive charge  and quark matter
has the opposite, all in accord with the conservation of electric charge.
The Coulomb force will tend to break regions of like charge into smaller
ones. The surface interface energy will resist. The competition is resolved
when the rare phase moves to lattice sites immersed in the dominant
phase. Charge is thus shielded over long ranges.
Fig.\ \ref{crystal_k300b180} shows how the geometric
phases vary in form from drops to rods to slabs, and in size and spacing
as a function of proportion of the quark phase.
The dimensions are typically nuclear
and the spacings vary widely as the boundaries of the mixed phase
region are approached. Of course nature interpolates
between these idealized geometries.

All of the above can be put into rigorous form. But I like the
physical motivation better. Of course it must be put into mathematical form
to preform an estimate of the crystalline structures and how they change
in size, spacing and form as the proportion changes at various
depths in the star \cite{glen95:c}.

\section{Pulsar Spin-down and Internal Structure}

Pulsars are rotating neutron stars with large magnetic fields. When the
field is not aligned with the rotation axis, the rotation of the field
stimulates electromagnetic radiation, sometimes over a very
broad band of frequencies, along a cone with the magnet
axis as center. A wind of electron-positron pairs is also created. 
These processes produce a torque on the pulsar. Rotational energy
is slowly radiated. 
The rates at which pulsar periods
change are in the
range $10^{-20}$ to $10^{-14}$ s/s.
It takes of the order of ten million years to radiate most of the
rotational energy.

As the pulsar spins down, it  evolves in shape from a spheroid to
 sphere. The interior density rises ever so slowly. As it rises
the density thresholds for various of the transformations that I have
discussed will be reached first at the center of the
star and then moving outward to embrace a slowly expanding region
of the star. We propose to use the timing structure
of pulsar spin-down to detect these transformations, especially the
deconfinement phase transition \cite{glent97:a}.
The coupling of rotation to the weak electromagnetic process of radiation
is favorable for detection, since the expansion of the transformed region
 typically has a time-scale of $10^5$ years.

The  energy loss equation representing processes of multipolarity $n$
is of the
form
 $$\frac{dE}{dt} =
   \frac{d}{dt}\Bigl(\frac{1}{2} I \Omega^2 \Bigr) =
      - C \Omega^{n+1}
$$
     where, for magnetic dipole radiation,
	    $ C=  \frac{2}{3} m^2 \sin^2 \alpha $,
	   $n=3$ ,
   $m$ is the magnetic dipole moment and $\alpha$ is the angle
   of inclination between magnetic moment and rotation axis.
   We shall refer to $n$ appearing in the energy-loss equation as the
 {\sl intrinsic} index.
 In the pulsar community, the above equation is integrated to yield
 $$ \dot{\Omega}=-K \Omega^n ~~~~(K=C/I)
 \,,~~~~~{\rm if~}I={\rm constant~or~}\Omega \ll 2I/I^\prime\,.
 $$
 and $n$ is known as the braking index. If the above equation held
 then the dimensionless measurable quantity
 $$\frac{\Omega \ddot{\Omega} }{\dot{\Omega}^2} =n
    $$
    would provide the value of the braking index, which for magnetic
    dipole radiation is 3, and which for the several pulsars for which
    it is measured has values around 2-2.5.
   The above two equations would hold rigorously for a {\sl rigidly}
   rotating magnetized body. 

However  a rotating star is not a rigid body. At high angular velocity
it is centrifugally deformed and becomes more spherical with time
as the star radiates.
Consequently the moment of inertia is not constant.
Moreover as the central density rises with decreasing angular velocity,
thresholds for various of the internal transformations spoken of above
will occur. Such changes in internal structure, which effect the distribution
of mass-energy in the star, bring about there own changes in the
moment of inertia. 
The equation that follows from the energy loss equation is therefore
$$ \dot{\Omega}= -\frac{C}{I(\Omega)}
  \biggl[1  + \frac{I^{\prime}(\Omega) \,
  \Omega}{2I(\Omega)}
   \biggr]^{-1} \Omega^n
       $$
and the measurable dimensionless braking index is not a constant
and not an integer
$$
  n(\Omega)\equiv\frac{\Omega \ddot{\Omega} }{\dot{\Omega}^2}
    = n
       - \frac{ 3  I^\prime \Omega +I^{\prime \prime} \Omega^2 }
	   {2I + I^\prime \Omega} \,.
$$
where $I^\prime = dI/d\Omega$.
We will find that for a first order phase transition, the observable index
will be far removed from its canonical value of $n$.

\section{Effect of Phase Transitions on Rotation}

The \eos  depends on the  composition
of matter, and hence the structure
of stars  depends on their composition. As we suggested
above, in the course of its spin-down, thresholds for various transitions
will be reached, first at the center and then in an expanding 
region. Figure \ref{omega_r_k300b180_d}
shows the evolution of the radial
boundaries between different phases (on the x-axis)
as the angular velocity decreases  with time.
We must incorporate these effects into a calculation of the
moment of inertia.

In classical mechanics  and Newtonian gravity
the moment of inertia involves a simple
and straightforward calculation. In \GR the calculation is quite
different because spacetime is warped, not only by the distribution
of mass-energy, but also by rotation. A particle dropped from great distance
from a point on the equatorial plane
onto a rotating star would not drop toward its center but would
acquire an ever increasing angular velocity in the same direction
as the angular velocity of the star. The local inertial frames are set 
into rotation. The angular velocity $\omega(r)$
\begin{figure}[tbh]
\vspace{-.5in}
\begin{center}
\leavevmode
\centerline{ \hbox{
\psfig{figure=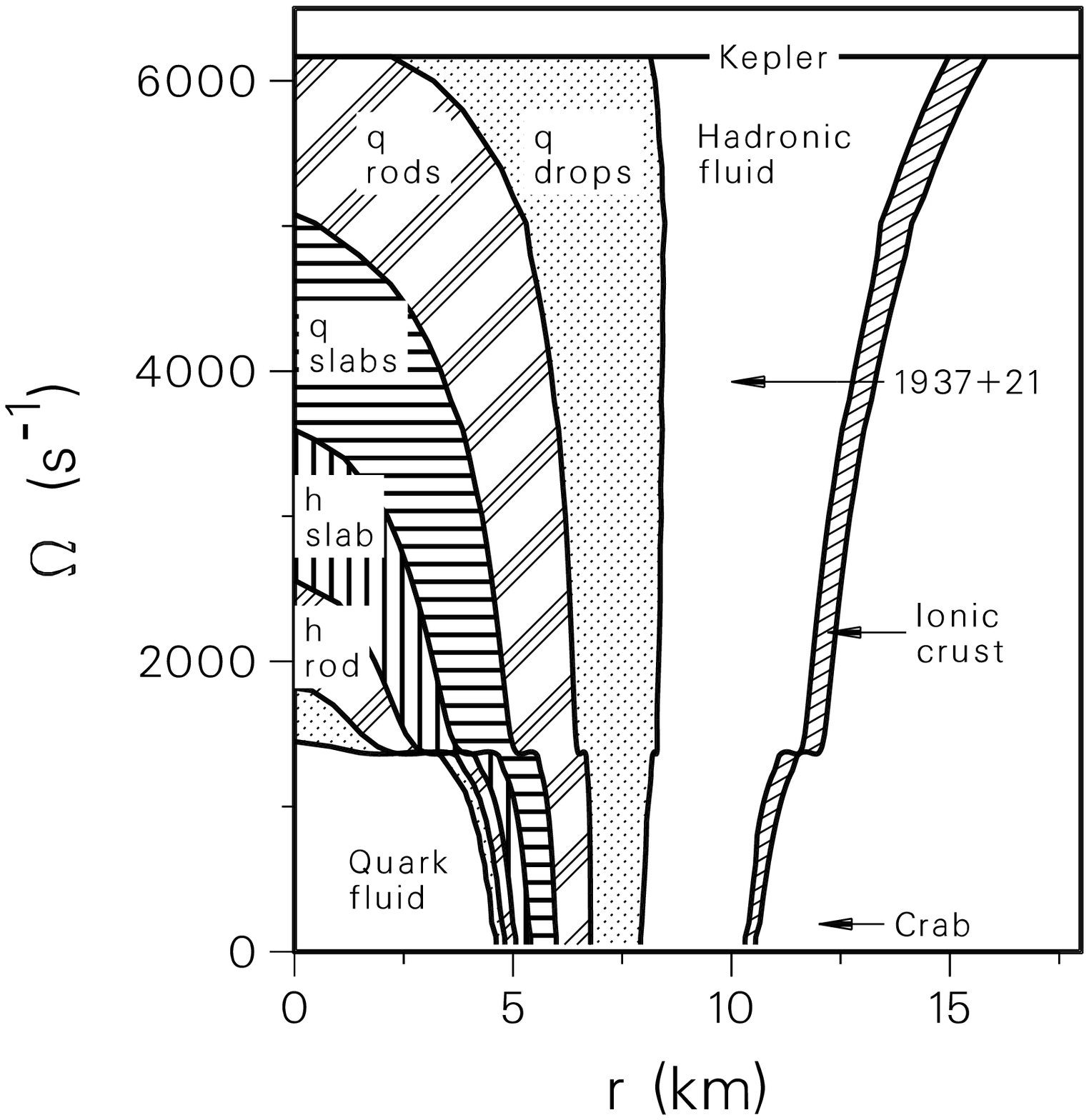,width=2.5in,height=3in}
\hspace{.6in}
\psfig{figure=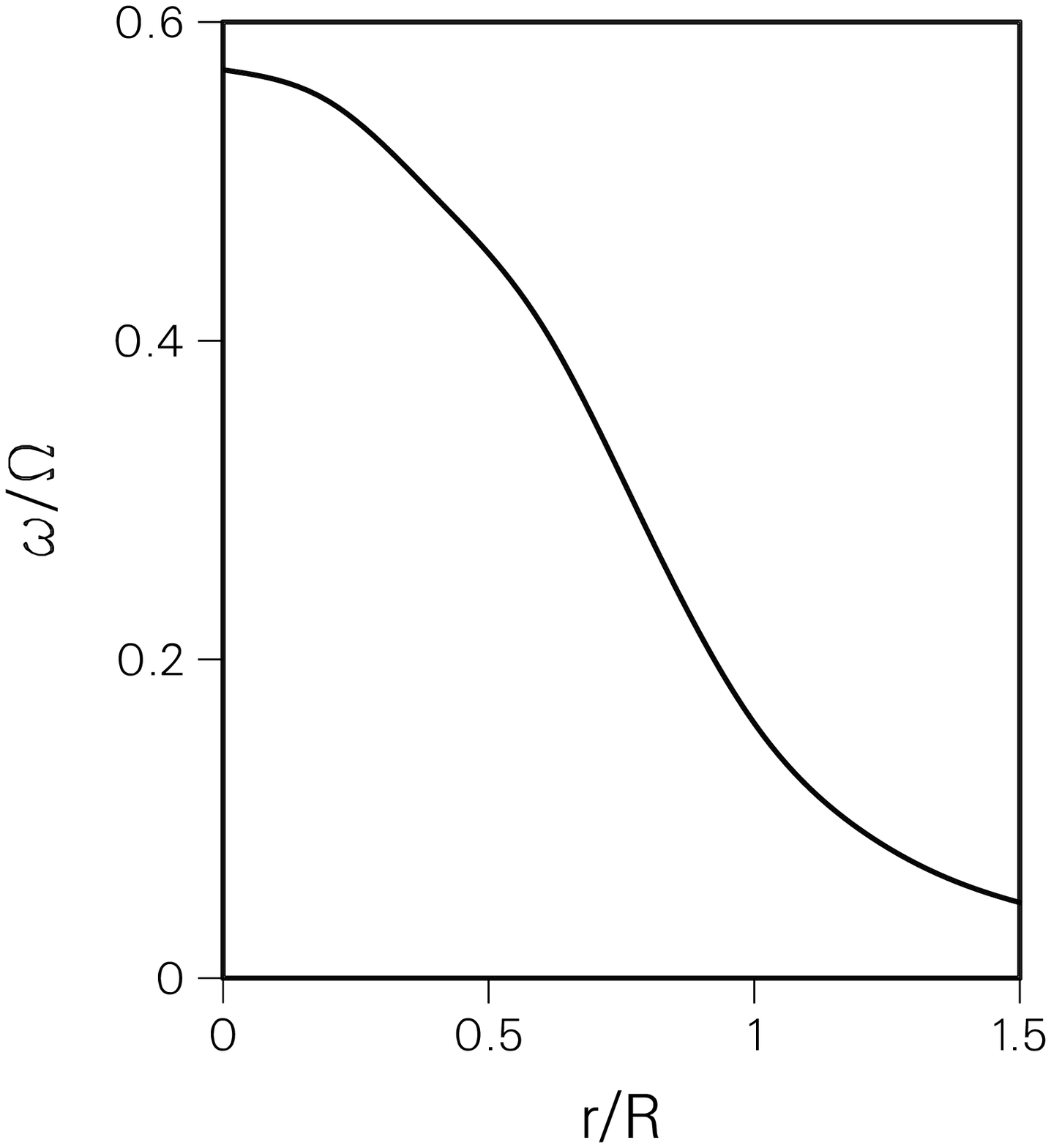,width=2.5in,height=3in}
}}
\begin{flushright}
\parbox[t]{2.7in} { \caption { \label{omega_r_k300b180_d} The radial
phase boundaries in a star having various angular velocities. Time
evolution is from the top.
}} \ \hspace{.4in} \
\parbox[t]{2.7in} { \caption { \label{drag} Frame dragging angular
velocity in units of the stellar angular velocity as a function
of distance from the center of the star \protect\cite{book} \courtesy.
}}
\end{flushright}
\end{center}
\end{figure}
of the local frames is a function
of distance from the star, being greatest at its center, and falling off
as $1/r^3$ outside the star. We show its ratio to that of the angular 
velocity of the star $\Omega$ in Fig.\ \ref{drag}. Of course this phenomenon 
effects the structure of the star itself. The reason is that the
centrifugal force acting on any fluid element in the star depends,
not on the angular velocity $\Omega$ but on the angular velocity
relative to that of the local inertial frames,
$\Omega-\omega(r)$. Prior to our work in 1992 there
was no expression available for calculation of the moment of inertia
except in the approximation that the star was not rotating. The approximation
is sometimes known as the slow approximation. The angular momentum
and moment of inertia are given by \cite{hartle67:a,hartle68:a}
$$ J_{\rm sph}=I_{\rm sph}\Omega = { {8\,\pi}\over 3} \; \int_0^R \, d r \,  r^4 \,
{ {\epsilon + P(\epsilon)}\over{ \sqrt{1 - 2\; m(r)\;   / \; r} } } \;
 [\Omega-\omega(r)] \;
 e^{-\Phi(r)} \;  \; $$
 where $e^{2\Phi(r)}$  is the time metric function $g_{tt}$
 of \OVp.
This is not adequate to our purpose because the distribution of mass, energy
and pressure in the star in the above expression refer to a nonrotating 
star. It lacks therefore the effect of the dependence on the internal 
structure  of the star---the way this structure changes with $\Omega$,
as well as the more elementary centrifugal distortion of the star. 
The above equation refers to a spherical \OV star.

We derived (in another connection) the expression for the moment of inertia
which incorporates the effects that are missing in the above result.
The expression depends, not on the \Sc metric, but on the metric
of a static rotating spacetime. The metrical functions are denoted by
$\lambda(r,\Omega)\cdots \psi(r,\Omega)$ and are given in 
Ref.\ \cite{glen92:b}
and refer to $g_{tt}=e^{2\nu},~g_{\phi t}=\omega(r)e^{2\psi}$, etc. 
The expression is
$$ J=I \Omega =
 4\pi  \int_0^{\pi/2} d\theta  \,
\int_0^{R(\theta)} dr\;
\frac{  e^{\lambda(r,\omega)} \; e^{\mu(r,\omega)} \; e^{\nu(r,\omega)}
   e^{\psi(r,\omega)}\;
       [\eps + P(\eps)]} {e^{2\nu(r,\omega) - 2\psi(r,\omega)}\; - \;
	   [ \Omega - \omega(r,\Omega)]^2 } \;
		[\Omega - \omega(r,\Omega)]\,. $$
The distribution of mass, energy density and pressure in this expression
refer now to those of a rotating star rather than a \OV star. 

Figure \ref{no2} shows how the moment of inertia changes with frequency
(and hence time). What is
\begin{figure}[tbh]
\vspace{-.5in}
\begin{center}
\leavevmode
\centerline{ \hbox{
\psfig{figure=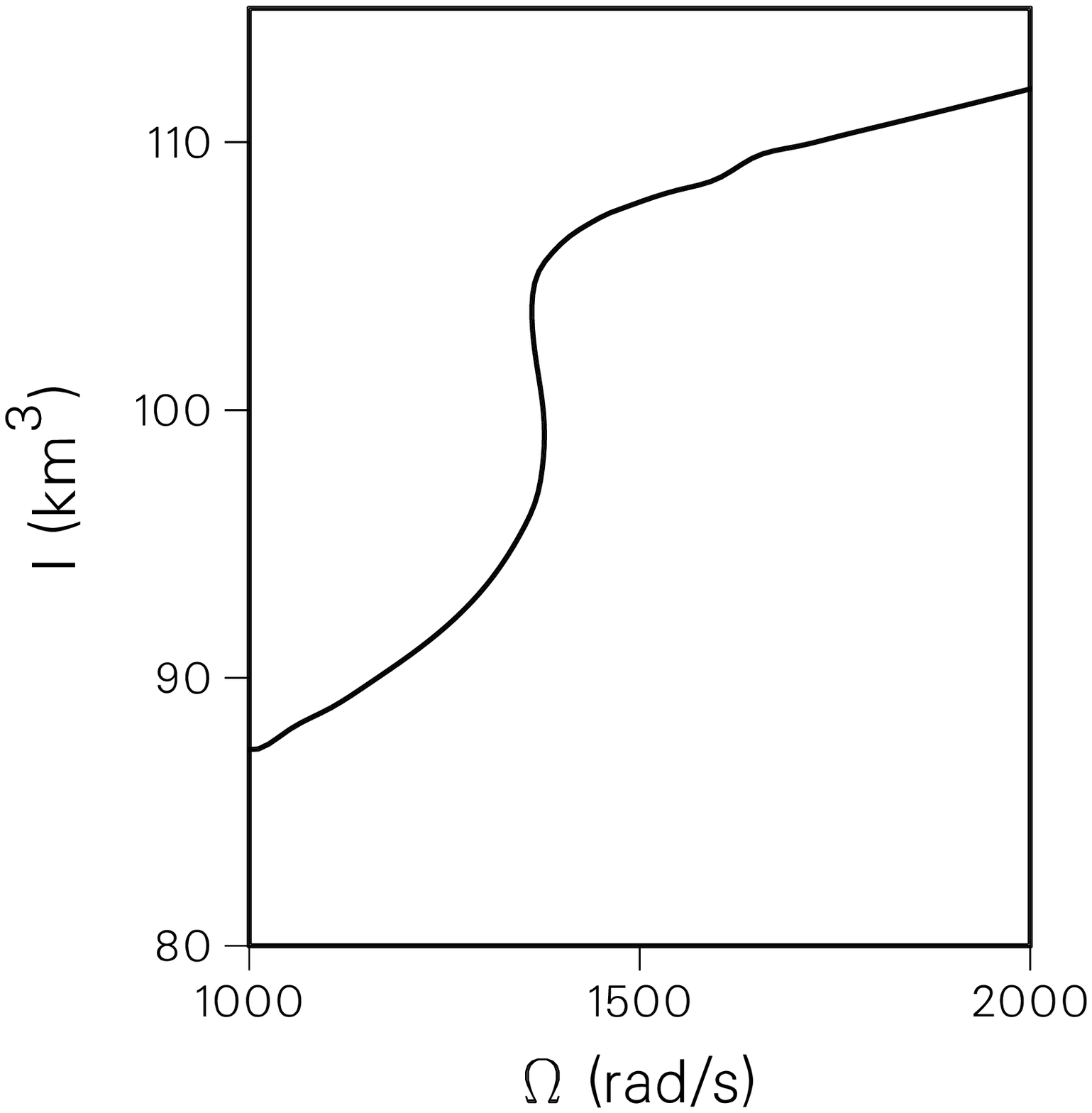,width=2.5in,height=3in}
\hspace{.6in}
\psfig{figure=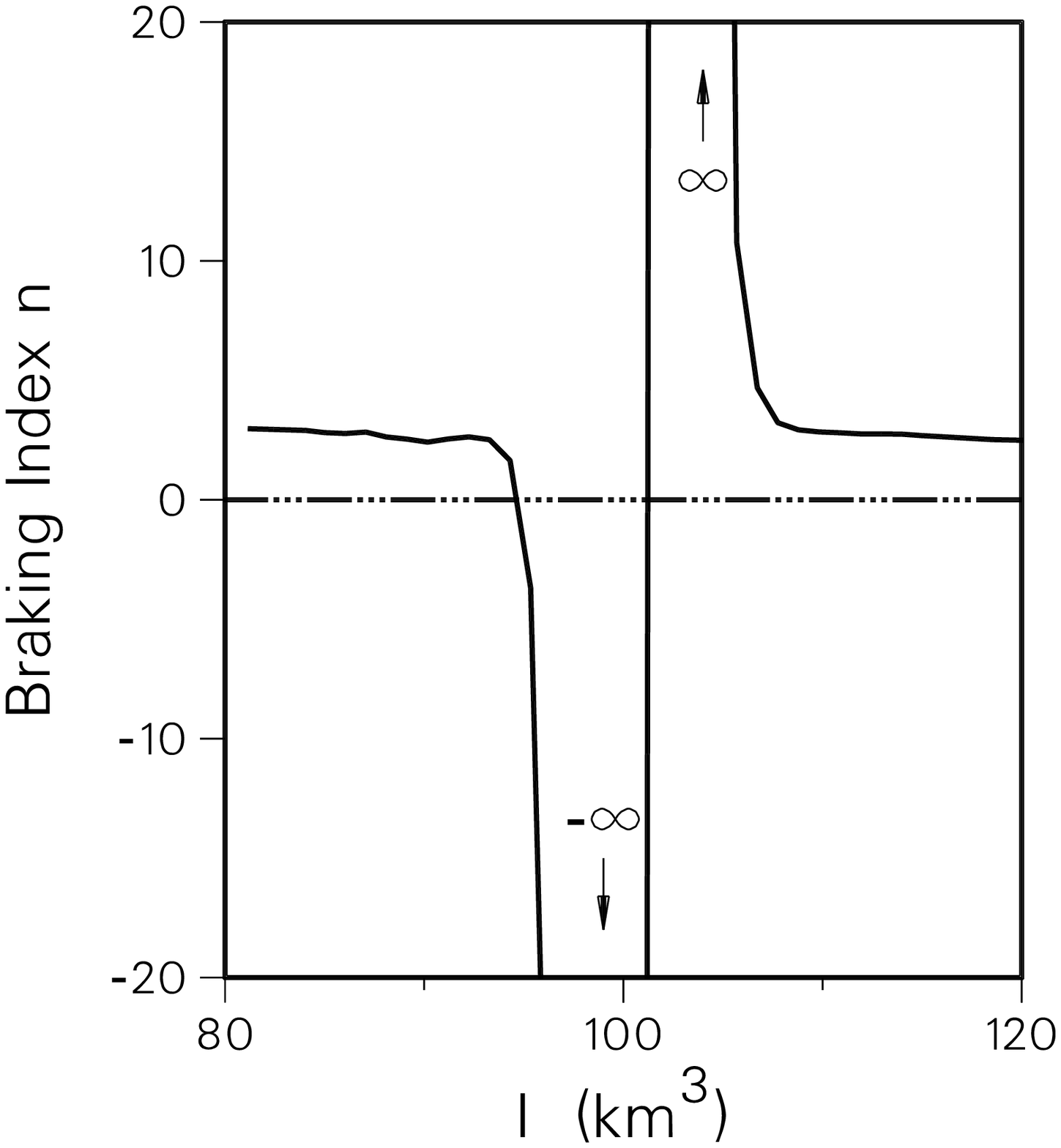,width=2.5in,height=3in}
}}
\begin{flushright}
\parbox[t]{2.7in} { \caption { \label{no2} Moment of inertia as a function
of rotational frequency. The ``bachbend'' occurs during the epoch
of phase transition. It is entirely analogous to backbending
observed in nuclei 
\protect\cite{glent97:a}.
}} \ \hspace{.4in} \
\parbox[t]{2.7in} { \caption { \label{ni} Measurable braking index.
The canonical value for magnetic dipole radiation is 3. The large
departures, which last for $\sim 10^5$ years, are caused by the growing
region in which nuclear matter is converted to quark matter 
\protect\cite{glent97:a}.
}}
\end{flushright}
\end{center}
\end{figure}
particularly noteworthy are: (1) As the frequency decreases from
high values, the moment of
inertia  decreases according to a particular trajectory, which is 
distinctly different when projected to low angular velocity from the 
trajectory actually followed. The two trajectories refer, so to speak
to a star of two distinct structures. From Fig.\ \ref{omega_r_k300b180_d}
we can see that the high frequency (ie lower density) structure corresponds
to a star with a mixed phase of confined and deconfined matter that
extends from the center of the star to about 8 km.
The lower frequency trajectory corresponds to a star (the same one but
later in time) that has a quark matter core of 4-5 km radius.
(2) During its evolution from one internal structure to the other, the
star actually spins up. The singularities and changes of sign in the
derivatives $I^\prime$ and $I^{\prime\prime}$ in the transition region
will produce a marked signal in the measurable braking
index $n(\Omega)$ as can be seen by the presence of these
derivatives in the index. In fact the dimensionless observable
$\Omega \ddot{\Omega} /\dot{\Omega}^2$ is singular at both 
turning points in Fig.\ \ref{no2} and is of opposite sign. The observable
braking index is shown in Fig.\ \ref{ni}. Its departure from the expected value
of three during the phase transition epoch is spectacular.  
If observed, it would be a clear signal of a first order phase transition.

We should enquire whether it is easy to measure an anomalous braking
index and how likely it is that an anomaly will be observed in the
population of some 700 pulsars presently known in our part of the
galaxy ($r<R_{\rm galaxy}/3$). 
Using the frequency interval
    $\Delta \Omega$ in which $n(\Omega)$
is greater than 6 or less than -3, and a typical
pulsar spin-down rate $\dot{\Omega}$
we find the duration of the transition epoch
 $$\Delta T \approx -\Delta \Omega/\dot{\Omega} \sim 10^5 ~{\rm years}$$
Since the mean life of pulsars
is $\sim 10^7$ years,
1/100 pulsars may exhibit the signal
(i.e. 7 of those 700 presently  known).
Moreover, although for normal pulsars $\ddot{\Omega}$ is very small and
difficult to measure, for a pulsar in the phase transition
epoch, this derivative is large---even infinite---at two
times during the  epoch. Consequently, the braking
index should be easy to measure for pulsars in the transition epoch.
In fact, difficulty in measuring the braking index could be used to
de-select candidates.

For the reasons discussed above, the prognosis for discovering first
order phase transitions in neutron stars appears to be excellent.
The signal is very strong, is easy to
measure and lasts for an appreciable fraction
of a pulsar's active lifetime, so that  roughly
ten percent of pulsars will be passing through a phase transition epoch, if
indeed the deconfinement transition occurs in neutron stars. 

\doe


 \end{document}